\documentclass[aps,prl,letterpaper,showpacs,superscriptaddress]{revtex4}
\usepackage{longtable}
\usepackage{graphicx}
\usepackage{epsfig}
\usepackage{amssymb}
\usepackage{color}

\newcommand{\beq}{\begin{equation}}
\newcommand{\eeq}{\end{equation}}
\newcommand{\bea}{\begin{eqnarray}}
\newcommand{\eea}{\end{eqnarray}}

\newcommand{\coon}{(Color online) }

\begin{document}

\title{Squeezed-field injection for gravitational wave interferometers}

\author{Henning Vahlbruch}
\author{Simon Chelkowski}
\author{Boris Hage}
\author{Alexander Franzen}
\author{Karsten Danzmann}
\author{Roman Schnabel}
\affiliation{Institut f\"ur Atom- und Molek\"ulphysik, Universit\"at Hannover and Max-Planck-Institut f\"ur Gravitationsphysik (Albert-Einstein-Institut), Callinstr. 38, 30167 Hannover, Germany}

\date{\today}

\begin{abstract}
In a recent table-top experiment we demonstrated the compatibility of three advanced interferometer techniques for gravitational wave detection, namely {\emph {power-recycling}},  {\emph {detuned signal-recycling}} and  {\emph {squeezed field injection}}. The interferometer's signal to noise ratio was improved by up to $2.8$\,dB beyond the coherent state's shot-noise. This value was mainly limited by optical losses on the squeezed field. We present a detailed analysis of the optical losses of in our experiment and provide an estimation of the possible nonclassical performance of a future squeezed field enhanced GEO\,600 detector.
\end{abstract}

\pacs{04.80.Nn, 07.60.Ly, 42.50.Dv}
\maketitle

\section{Introduction}

The injection of a squeezed field into the dark port of a gravitational wave (GW) Michelson interferometer was first proposed 25 years ago by C. Caves \cite{Cav81}. The goal of this early proposal was the reduction of the measurement's shot-noise.
Later in the 1980s it was realized that squeezed states can also be used to reduce the overall quantum noise in interferometers including radiation pressure noise, thereby beating the standard-quantum-limit (SQL) \cite{Unruh82,JRe90}.
Independently from squeezed-field injection other advanced interferometer techniques have been developed to reduce the quantum noise. \emph{Arm cavities} as well as \emph{power-recycling} \cite{DHKHFMW83pr} and \emph{signal-recycling} \cite{Mee88} have been invented to improve the interferometer's signal to shot-noise ratio \cite{geo04}. In the first two cases additional mirrors form cavities which are resonant for the carrier laser light. In the latter case an additional mirror is placed into the interferometer's dark signal port establishing a carrier detuned but signal resonant cavity.
Recently it was discovered that at radiation pressure dominated frequencies signal-recycling can also be used to beat the SQL \cite{BCh01a}. However, the compatibility of recycling techniques and squeezed-field injection is an important question which has been investigated theoretically:
Gea-Banacloche and Leuchs showed that the techniques of power-recycling and squeezed-field injection are fully compatible \cite{GLe87},
Chickarmane\,{\it et\,al.}\,\cite{CDh96} found compatibility of signal recycling and squeezed-field injection for the shot-noise limited regime.
Furthermore the analysis by Harms\,{\it et\,al.} \cite{HCCFVDS03} showed that the same is true for detuned signal-recycling at shot-noise as well as radiation pressure noise dominated frequencies  thereby proposing that all the techniques can simultaneously be used to reduce quantum noise in interferometers. It is therefore very likely that future GW interferometers will use a combination of all those techniques.
So far only a few squeezed light enhanced interferometers have been demonstrated experimentally, e.g.\,table-top Mach-Zehnder and polarization interferometers \cite{XWK87,GSYL87}, respectively.
Recently a squeezing enhanced power-recycled Michelson interferometer has been reported already bearing more resemblance to a GW detector \cite{KSMBL02}.

In this paper we present a detailed analysis of the optical loss budget for the squeezed-field in interferometers with power-recycling,  detuned signal-recycling and squeezed-field injection. The combination of all three techniques has been demonstrated in a table-top-experiment \cite{VCHFDS05}. The interferometer's signal to noise ratio was improved by up to $2.8$\,dB (in power) beyond the coherent state's shot-noise; this value was mainly limited by optical loss. We provide an estimation of the possible nonclassical performance of a future squeezed-field enhanced GEO\,600 detector reaching a sensitivity of 6\,dB better than its photon shot-noise.

\section{Experimental}

\begin{figure*}[t!]
\hspace{0mm}
\includegraphics[width=15cm]{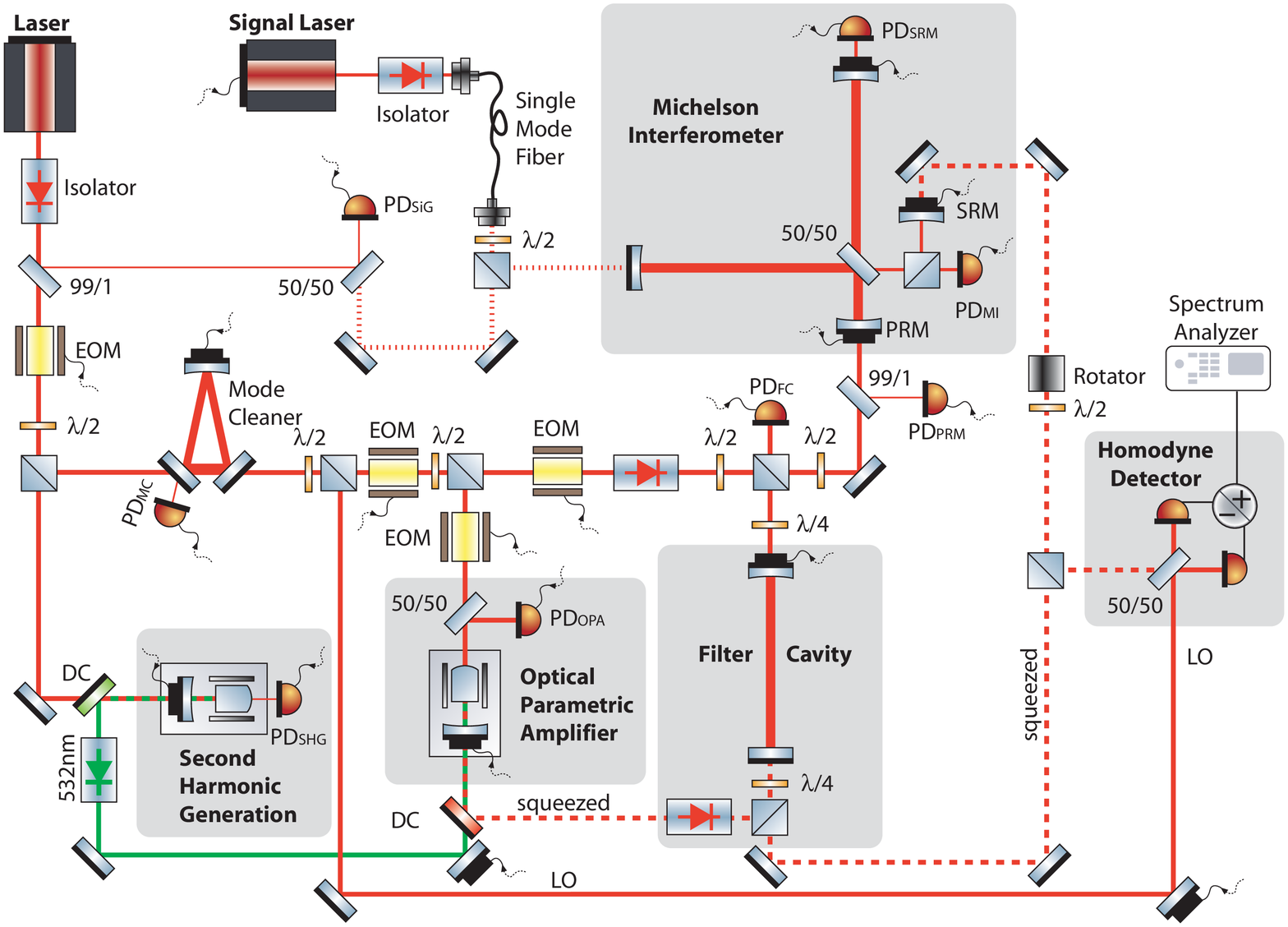}
  \vspace{0mm}
\caption{\coon Schematic of the experiment. Amplitude
squeezed light is generated in an OPA cavity of controlled length.
The detuned filter cavity provides frequency dependent squeezing
suitable for a broadband quantum noise reduction of a shot-noise
limited dual-recycled Michelson interferometer. SHG: second harmonic generation; OPA: optical
parametric amplifier; EOM: electro optical modulator; DC: dichroic mirror; LO: local oscillator;
PD: photo diode; PRM: power recycling mirror; SRM: signal recycling mirror;
\vrule width 3mm depth -.6mm\,: piezo-electric transducer.}
  \label{experiment}
\end{figure*}

Fig.~\ref{experiment} shows the optical layout of our experiment. Frequency dependent squeezed light was generated and injected into the dark port of a power- and signal-recycled Michelson interferometer. Both recycling cavities had lengths of about 1.21\,m and the reflectivities of the power recycling mirror (PRM) and the signal recycling mirror (SRM) were both 90\,\%.
The main laser source was a monolithic non-planar Nd:YAG ring laser of 2\,W single mode output power operating at 1064\,nm. About 50\,\% of the laser power was used for second harmonic generation (SHG) to produce the pump field for the optical parametric amplifier (OPA).
The remainder was transmitted through a mode cleaner ring cavity to provide a spectrally and spatially filtered beam in the TEM$_{00}$ spatial mode. This beam was split and used for locking the filter cavity, as a local oscillator (LO) for homodyne detection, for the Michelson interferometer and as a seed beam for the OPA. The OPA was constructed from a type I phasematched MgO:LiNbO$_3$ crystal inside a hemilithic resonator \cite{CVHFLDS05}. The resonator was formed by a high-reflection (HR) coating (R=99.996\,\%) of the crystal surface on one side and an externally mounted cavity mirror with a reflectivity R=96.7\,\%. The OPA was seeded through the HR-surface and its length servo controlled.
The pump field was injected through the coupler, double-passed the nonlinear crystal and provided a classical gain of 5. Locking the OPA to deamplification generated a broadband amplitude quadrature squeezed beam of about 92\,$\mu$W at 1064\,nm. This beam was first sent through an isolator and then used for a broadband nonclasssical sensitivity improvement of the Michelson interferometer. The latter one was stabilized on a dark fringe and its signal-recycling cavity locked to $+10$\,MHz sideband frequency. The squeezed-field was reflected from a filter cavity which was locked to $-10$\,MHz providing the optimum frequency dependent orientation of the squeezing ellipse. Such a scheme was proposed in \cite{KLMTV01} and first demonstrated in \cite{CVHFLDS05}.
The squeezed beam from the filter cavity was then injected into the signal recycling cavity (SRC) passing a Faraday rotator. This gave spatial degeneracy between the reflected squeezing and the signal output beam of the interferometer as shown in Fig.~\ref{experiment}. The combined field was guided to a homodyne detector that was built from two electronically and optically matched photodetectors based on Epitaxx ETX1000 photodiodes.
In order to be able to measure the improvement of the interferometer's signal to noise ratio we injected a single sideband modulation field, similar to the approach in \cite{GdeVine}. The signal was generated utilizing a second monolithic non-planar Nd:YAG ring
laser which was frequency locked to the main laser with tunable beat frequency in the range of 5--15\,MHz between both light sources and injected through one of the interferometer end mirrors. Note that for controlling the complete experiment in all its degrees of freedom, for example cavity-lengths and crystal temperatures, as many as 16 locking loops came into operation. A more detailed description can be found in \cite{VCHFDS05} and \cite{CVHFLDS05}.

\begin{figure*}[t!]
\hspace{-70mm}
\includegraphics[width=9cm]{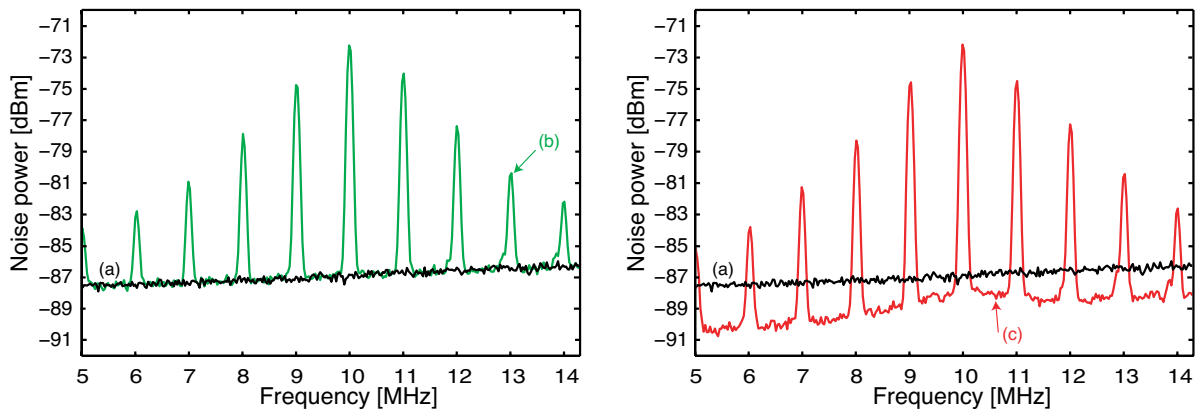}
\caption{\coon Amplitude quadrature power spectra of the dual-recycled Michelson interferometer with and without nonclassical noise reduction. The interferometer was stably locked with $+10$\,MHz detuned signal-recycling cavity. Left: (a) Shot-noise measured with a blocked signal beam at the homodyne detector; (b) Shot-noise limited signals; Right: (c) broadband squeezing enhanced signal-to-noise ratios of up to 2.8\,dB below shot-noise. Both signal curves are a sequential composition of ten individual measurements; in each measurement  a single signal of constant height was injected.
All spectra were analyzed with 100\,kHz  resolution bandwidth and 100\,Hz video bandwidth, averaging over five subsequent measurements and were at least 5\,dB above the detection dark noise which was taken into account.}
\label{signals}
\end{figure*}

\section{Results and discussion}

\begin{figure*}[t]
\centerline{\includegraphics[width=13cm]{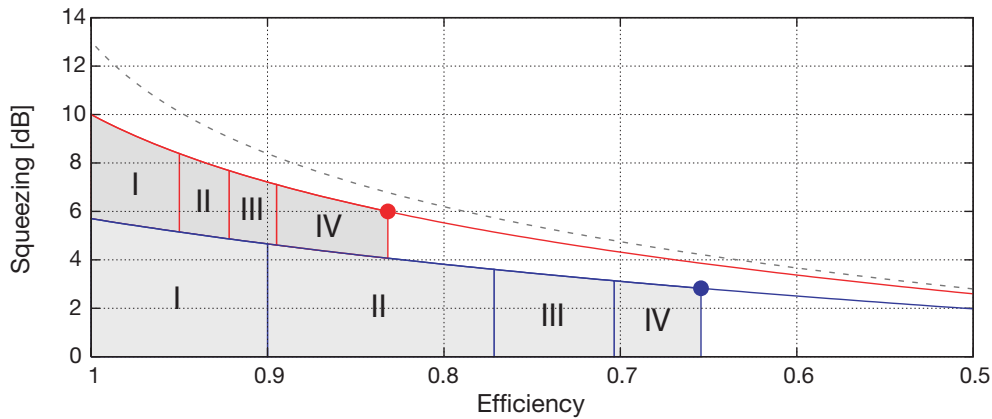}}\vspace{-1mm}
\caption{\coon Observed squeezing versus detection efficiency for
three different examples. The lower curve presents the influence
of the optical loss budget in our experiment. The upper solid line
presents the envisaged improved situation for a squeezed-field
enhanced GEO\,600 detector. 6\,dB nonclassical noise supression is
feasable if one starts with 10\,dB squeezing allowing 17\,\%
overall loss. The upper dashed line represents the situation if a
squeezed light source generates 13\,dB squeezing. The shaded areas
correspond to escape efficiency (I) and losses due to
modematchings (II), isolators/rotators (III) and Photodiodes
(IV).} \label{loss}
\end{figure*}

Fig.~\ref{signals} shows the nonclassical signal to noise improvement in our experiment. The peaks are injected single sideband modulation signals. Their different heights correspond to the signal transfer function of the signal-recycling cavity (SRC). The shot-noise reference is given by the flat lines. In the right picture a broadband squeezing enhanced signal-to-noise ratio of up to 2.8\,dB is observed. This value is mainly limited by losses on the squeezed-field. A clear goal for future applications in GW interferometers is to reduce the overall loss budget to a minimal possible value. Let us first discuss the loss budget in our table-top experiment.
First of all loss in terms of absorption and scattering inside the OPA cavity reduces the squeezing; this is described by the OPA's escape efficiency which is given by the ratio of output coupling decay rate to the total cavity decay rate. Absorption and scattering of the crystal and its anti-reflection coating contributed to a round-trip loss by about (0.37$\pm$0.05)\,\% resulting in an escape efficiency of 90\,\%.
Another major loss source (6\,\%) was the Faraday isolator (Linos, FI-1060-5 SI) that protected the OPA from most backscattered light.
The squeezed beam was then mode matched into the filter cavity with an efficiency of 95\,\%. The technique of squeezed-field injection required double-passing a Faraday rotator (Linos, FR-660/1100-5 SI). The double-pass transmission was measured to be 97\,\%. The mode matching efficiency of the squeezed-field into the SRC was 95\,\%. Finally the squeezed-field sensed loss in the homodyne detector due to the non perfect mode matching with the local oscillator of efficiency 95\,\% and the Epitaxx ETX1000 photodiodes with 93\,\% quantum efficiency.
The overall loss budget limited our squeezed-field detection efficiency to 65\,\%. The result was a detected nonclassical noise supression of 2.8\,dB at 5\,MHz. From these numbers one can calculate that about 5.7\,dB squeezing was generated inside the OPA.
The relation of these numbers are also given in the lower curve in Fig.~\ref{loss}.
The poorer squeezing of about 2.0\,dB at 14\,MHz in Fig.~\ref{signals} was due to the limited bandwidth of the OPA of 20\,MHz which will not be an issue for at the acoustic band for GW detectors.
The further reduced squeezing at 10\,MHz is due to loss inside the signal-recycling cavity.

In the following paragraph we estimate the possible nonclassical performance of a future squeezed-field enhanced GEO\,600 detector at sideband frequencies at which its sensitivity is shot-noise limited, typically above a few hundred Hertz, as envisaged in \cite{SHSD04}.
After applying classical noise suppression to our squeezed light source to enable squeezing in the GW band as demonstrated in \cite{MGBWGML04} the optical layout of our demonstration experiment can directly be applied to improve the sensitivity of large scale signal-recycled interferometers.
The key issue limiting the amount of squeezing that can be employed in GW detectors is the availability of high quality optics. Assuming the possibility to generate 10\,dB squeezing \emph{inside} the OPA within the detection bandwidth of the GW detector, only losses of 17\,\% are arguable to end up with a 6\,dB improved signal to noise ratio.
A nonclassical  noise suppression of 7\,dB has already been obeserved for GW detector laser wavelength of 1064\,nm.
Counting back the existing detection losses and the outcoupling efficiency of the OPO one concludes that indeed at least 10\,dB of squeezing has been generated inside a monolithic OPO \cite{LRBCBG99}. In that experiment no additional cavities degraded the squeezing and the escape efficiency was intrinsically higher due to the absence of intra-cavity surfaces in a monolithic design. However, for applications in interferometry the length of the OPO or OPA has to be locked to the main laser source which requires a tunable cavity length. One possibility to prevent the amount of squeezing is to use an OPA design analog to the one used in this work but with a higher escape efficiency, for example of 95\,\%. This can be done by lowering the output coupler reflectivity to 93\,\%. In return the green pump power has to be increased quadratic with respect to the outcoupling mirror transmissivity.

Further loss in our experiment was caused by the Faraday isolator and rotator. Both devices were built from Terbium Gallium Garnet with absorption of about 0.5\,\%\,cm$^{-1}$. The main differences between the isolator and the rotator used were the number and quality of polarizing beam splitters and, in the latter case, the possibility to tune the magnetic field. With optimized magnetic field the measured double pass loss of the rotator was just 3\,\%. This value was mainly limited by the non perfect anti-reflection coatings of the crystal itself and the quality of the (single) polarizing beam splitter (PBS). Further optimization of coatings and PBS quality should provide an entire loss for the required three passes through isolators/rotators of all together 3\,\% only.
Another sensitive item with regard to induced losses are the mode matching efficiencies. At the beginning of the chain the generated squeezing has to be injected into the filter cavity. In case of GEO\,600 the (suspended) filter cavity might have a length of 600\,m. Compared to our table top experiment the mode matching efficiency should increase due to the large Rayleigh range of the beam. The cavity mode itself should be pure Gaussian since no intra-cavity optics and no high power laser beam deteroriates the beam shape. However, large mode matching optics and couplers of high homogeneity are required. From the experience with current GW detectors a mode matching efficiency of 99\,\% should be achievable.
Probably a bit more critical is the squeezed-field injection into the signal-recycling cavity. Due to the high circulating light power inside the interferometer thermal lensing effects at the beam splitter can lower the Gaussian beam profile quality. One might expect a mode matching efficiency of 98\,\% but since thermal compensation schemes \cite{WGEO04} are likely to be applied this value should also reach 99\,\%.
As shown in Fig.~\ref{signals} optical loss inside the signal recycling cavity for example in the Michelson beamsplitter effects the amount of squeezing. However this is a frequency dependent loss source since the squeezing gets reduced most exactly at the SRC-detuning frequency. But also in case of non-negligible losses inside this cavity only the performance at the optical resonance is degraded from the 6\,dB goal and the performance at neighbouring detection frequencies is then unaffected providing an improved detection bandwidth. Refering to the homodyne-readout scheme planned for the advanced detectors the signal-beam from the interferometer has to be modematched on a 50/50 beamsplitter with an intense local oscillator. Since a beam might be picked from the anti-reflection coating of the Michelson beamsplitter the beam shape should be close to the interferometer's signal beam. In all probability again a mode matching efficiency of 99\,\% should be achievable. At last but of particular importance the photodetectors themselves have to be mentioned. Assuming losses as described so far the quantum efficiency of the photodetectors  must be at least as high as 93\,\%. This value is associated with the best currently available photodiodes usable in our table-top experiment.
Here we assume that the same high quantum efficiency can be achieved for high power (up to one watt) photodiodes which are required in GW interferometers.
The noise budget envisaged here provides an overall detection efficiency of the squeezed-field of 83\,\%. The upper solid line in Fig.~\ref{loss} shows that for initally 10\,dB squeezing still 6\,dB noise suppression is available for improving the noise spectral density of a GW detector by a factor of 4 in power.

\section{Conclusion}

We have analyzed the noise budget on the squeezed-field in our squeezing enhanced power- and signal-recycled Michelson interferometer including filter cavity for the generation of frequency dependent squeezing. It seems to be feasible with current technology to increase the overall detection efficiency from 65\,\% in our table-top experiment to 83\,\% when scaled to a GW detector like GEO\,600. From previous experiments one might infere that the goal of a nonclassical sensitivity improvement of 6\,dB is possible.

 This work has been supported by the Deutsche Forschungsgemeinschaft and is part of Sonderforschungsbereich 407.


\section{References}

\begin{thebibliography}{12}

\bibitem{Cav81} Caves C M, Phys. Rev. D {\bf 23}, 1693 (1981).

\bibitem{Unruh82} Unruh W G, in  {\it Quantum Optics, Experimental Gravitation, and Measurement Theory}, edited by P.~Meystre and M.~O.~Scully (Plenum, New York, 1983), p.~647--660.

\bibitem{JRe90} Jaekel M T and Reynaud S, Europhys.~Lett.~{\bf 13}, 301 (1990).


\bibitem{DHKHFMW83pr} Drever R W P {\it et al}\ in {\it Quantum Optics, Experimental Gravitation, and Measurement Theory}, edited by P.~Meystre and M.~O.~Scully (Plenum, New York, 1983), p.~503--514.

\bibitem{Mee88} Meers B J, Phys. Rev. D {\bf 38}, 2317 (1988).

\bibitem{geo04} Abbott B \emph{et al}, Nuclear Instruments and Methods in Physics Research Section A \textbf{517}, 154 (2004).

\bibitem{BCh01a} Buonanno A and Chen Y, Class. Quantum Grav. {\bf 18}, L95 (2001).

\bibitem{GLe87} Gea-Banacloche J and Leuchs G, J. Mod. Opt. {\bf 34}, 793 (1987).

\bibitem{CDh96} Chickarmane V and Dhurandhar S V, Phys. Rev. A {\bf 54}, 786 (1996).

\bibitem{HCCFVDS03}Harms J, Chen Y, Chelkowski S, Franzen A, Vahlbruch H, Danzmann K and Schnabel R, Phys. Rev. D {\bf 68}, 042001 (2003).

\bibitem{XWK87} Xiao M, Wu L A and Kimble H J, Phys. Rev. Lett. {\bf 59}, 278 (1987).

\bibitem{GSYL87} Grangier P, Slusher R E, Yurke B and LaPorta A, Phys. Rev. Lett. {\bf 59}, 2153 (1987).

\bibitem{KSMBL02} McKenzie K, Shaddock D A, McClelland D E, Buchler B C and Lam P K, Phys. Rev. Lett. {\bf 88}, 231102 (2002).

\bibitem{VCHFDS05} Vahlbruch H, Chelkowski S, Hage B, Franzen A, Danzmann K and Schnabel R, Phys. Rev. Lett. {\bf 95}, 211102 (2005)).


\bibitem{CVHFLDS05} Chelkowski S, Vahlbruch H, Hage B, Franzen A, Lastzka N, Danzmann K and Schnabel R, Phys. Rev. A {\bf 71}, 013806 (2005).

\bibitem{KLMTV01} Kimble H J, Levin Y, Matsko A B, Thorne K S and Vyatchanin S P, Phys. Rev. D {\bf 65}, 022002 (2001).

\bibitem{GdeVine} de~Vine G, Shaddock D A, McClelland D E, Class. Quantum Grav. {\bf19}, 1561Ê(2002).


\bibitem{SHSD04} Schnabel R, Harms J, Strain K A and Danzmann K, Class. Quantum Grav. \textbf{21}, S1045 (2004).


\bibitem{MGBWGML04} McKenzie K, Grosse N, Bowen W P, Whitcomb S E, Gray M B, McClelland D E and Lam P K, Phys. Rev. Lett. {\bf 93}, 161105 (2004).

\bibitem{LRBCBG99} Lam P K, Ralph T C, Buchler B C, McClelland D E, Bachor H A and Gao J, J. Opt B: Quantum Semiclass. Opt.1 (1999) 469-474.

\bibitem{WGEO04} Willke B \emph{et al}, Class. Quantum Grav. {\bf 21}, S417 (2004).



\end{thebibliography}
\end{document}